\magnification=\magstep1
\tolerance 500
\rightline{TAUP-2490-98}
\rightline{22 April, 1998}
\centerline{\bf Second Quantization of the Stueckelberg }
\centerline{{\bf Relativistic Quantum Theory and Associated
Gauge Fields}
\footnote{*}{ Presented at the First International Conference on
Parametrized Relativistic Quantum Theory, PRQT '98, Houston,
Feb. 9-11, 1998.}}
\centerline{L.P. Horwitz\footnote{\dag} {Also at Department of
Physics, Bar Ilan University, Ramat Gan, Israel}}
\centerline{School of Physics}
\centerline{Raymond and Beverly Sackler Faculty of Exact Sciences}
\centerline{Tel Aviv University, Ramat Aviv, Israel}
 \bigskip
\noindent
{\it Abstract\/} The gauge compensation fields induced by the
differential operators of the Stueckelberg-Schr\"odinger equation are
discussed, as well as the relation between these fields and the
standard Maxwell fields.  An action is constructed and the second
quantization of the fields carried out using a constraint
procedure. Some remarks are made on the properties of the second
quantized matter fields.
\bigskip
\noindent
{\bf I. Introduction.}
\smallskip
\par There has been considerable progress in recent years in the study
of the relativistic quantum theory proposed (for a single particle) by
Stueckelberg$^1$ and generalized by Horwitz and Piron$^2$ to many body
systems.  Collins and Fanchi$^3$ arrived at this structure by
considering the relativistic current conservation law, and Fanchi$^4$
has made many studies of the properties of the theory.  In particular,
he has shown that there is no Klein paradox.$^5$ It has been
shown$^2$ that the Newton-Wigner operator,$^6$  originally constructed
for on-shell wave equations, emerges from the Stueckelberg theory,
which is intrinsically off-shell, as the operator
$$ {\bf x}_{op} = {\bf x} - {1 \over 2} \{t,{ {\bf p} \over E} \},
 \eqno(1.1)$$
in a direct integral representation
over masses, at each mass value.  The Landau-Peierls$^7$ uncertainty
relation, $\Delta p \Delta t \geq \hbar/c$ has also been shown to
arise as an exact mathematical bound$^8$ for the dispersion of the
operator
$$ t_{op} = t - {1 \over 2} \{x, {E \over p} \}, \eqno(1.2)$$
which has an evident semiclassical interpretation.
\par The two body quantum relativistic bound state problem for
spinless particles with invariant action-at-a-distance potentials has
been solved$^9$; the differential equation for the radial part of the
wave function (for which the variable is the spacelike invariant
separation $\rho = \sqrt{x_1 - x_2)^2 -(t_1 - t_2)^2 }$) is identical
to the non-relativistic radial Schr\"odinger equation with potential
of the same functional form.  The eigenvalues of the reduced motion
Hamiltonian therefore coincide with the Schr\"odinger energy spectrum,
but correspond to mass shifts induced by the interaction.  The
observed center of mass energy of the system is given by (see ref. 9
for further discussion)
$$ E_n = \sqrt{M^2 + 2Mk_n}, \eqno(1.3)$$
where $k_n$ is the (Schr\"odinger) spectrum of the reduced motion, and
$M$ is a dimensional scale parameter
 corresponding to the sum of the
Galiean limiting masses of the two particles of the system.  For small
excitations, $E_n \cong M + k_n$, up to terms of higher order in $1/c$
(relativistic corrections).
\par I wish to review here the general gauge structure of the theory,
the relation of the gauge fields generated by the
Stueckelberg-Schr\"odinger equation to the Maxwell fields,
 and their second quantization
based on a constraint formalism$^{10.11}$.
\bigskip
\noindent
{\bf II. The Gauge Fields.}
\par I begin with the statement of the Stueckelberg-Schr\"odinger
equation for the wave function representing a free one-particle state,
$$ i {\partial \over \partial \tau}\psi_\tau(x) = {p_\mu p^\mu \over
2M}  \psi_\tau(x), \eqno(2.1)$$
where $x \equiv x^\mu = ({\bf x},t)$, and $\tau$ is the invariant
parameter of evolution.  The operator $p^\mu$ is realized by $-i\partial
/\partial x_\mu$ in this representation.
 We recognize that, in addition to the
coordinate ${\bf x}$, $t$ is also an observable in the quantum theory,
for which there is a self-adjoint time operator.  This structure is
necesssary for the manifest covariance of the quantum theory for
which, under the action the Lorentz group, the space and time
variables transform (linearly) among each other.\footnote{*}{ Note
that the parameters of the Lorentz group are not associated with the
position or velocity of a particle.  The association is often made
classically when one describes the particle in terms of a motion
induced on a particle at rest by transforming to a moving frame.  The
acceleration of a particle due to forces cannot be accounted for in
this way, since this would involve transformation to a non-inertial
frame, going beyond the applicability of special
relativity. Accelerated motions of the particle are accounted for in
the Stueckelberg theory as a result of covariant dynamical equations.}
 The standard relativistic quantum field theories admit the Lorentz
transformation in a consistent way by interpreting both ${\bf x}$ and
$t$ as parameters.  As we shall see, the second quantization of the
Stueckelberg theory has the same property, but includes the evolution
parameter $\tau$ for the description of dynamical processes (it does
not necessarily correspond to a label for spacelike surfaces, as in the
formulation of Schwinger and Tomonaga$^{12}$, or to the proper time of
any system). The coordinate representations generated by these fields are over
space-time, as for the one-body quantum theory of Stueckelberg.$^1$
\par   If we admit a local gauge transformation
$$ \psi_\tau'(x) = e^{i\Lambda(x,\tau)} \psi_\tau(x), \eqno(2.2)$$
the equation $(2.1)$ must be altered in order to admit gauge
covariance by the addition of gauge compensation fields.$^{13}$ The
modified equation reads
$$ \bigl( i {\partial \over \partial\tau} + e_0 a_5 \bigr)
\psi_\tau(x) = {(p^\mu -e_0 a^\mu)(p_\mu - e_0 a_\mu) \over 2M}
\psi_\tau(x), \eqno(2.3) $$
 which is clearly gauge covariant if
$$\eqalign{{a^\mu}'  &= a^\mu + { 1\over e_0} \partial^\mu \Lambda \cr
a_5 ' &= a_5 + {1 \over e_0} \partial_5\Lambda, \cr} \eqno(2.4)$$
which we can write in terms of a five-component field ($\alpha =
0,1,2,3,5; \, \partial_5  \equiv \partial_\tau $)
$$ a_\alpha ' = a_\alpha + {1 \over e_0} \partial_\alpha \Lambda.
\eqno(2.5)$$
The field strengths
$$ f_{\alpha \beta} = \partial_\alpha a_\beta - \partial_\beta
a_\alpha \eqno(2.6)$$
are gauge invariant, and field equations of second order can be
generated if the term $f_{\alpha \beta}f^{\alpha \beta}$ is present in
the Lagrangian.  In order to determine the coefficients in the model
Lagrangian that we shall write down, we shall need some information
about the conserved current associated with Eq. $(2.3)$.
\par Differentiating the probability density $\rho_\tau (x)  = \vert
\psi_\tau(x) \vert^2 $ with respect to $\tau$, and using Eq. $(2.3)$,
one finds that
$$ \partial_\mu  j^\mu + {\partial \rho \over \partial \tau} = 0,
\eqno(2.7)$$
where
$$ j^\mu_\tau (x) = {1 \over 2Mi} \{ \psi_\tau^* (\partial ^\mu -
ie_0 a^\mu) \psi_\tau - [(\partial^\mu - ie_0a^\mu) \psi_\tau]^*
\psi_\tau \}. \eqno(2.8)$$
Since $\rho_{\pm \infty} (x) = 0$ $^{1, 14,15}$ (pointwise), it follows from
$(2.7)$ that  the integrated current
$$ J^\mu = \int_{-\infty}^\infty d\tau \, j^\mu_\tau (x)\eqno(2.9)$$
satisfies
$$ \partial_\mu J^\mu (x) = 0, \eqno(2.10)$$
and may therefore be identified with the Maxwell current.
Writing a Lagrangian density which generates the
Stueckelberg-Schr\"odinger equation as a field equation, along with
the second order equations for the gauge compensation fields, one
finds that these field equations have the form
$$ \partial_\beta f^{\alpha \beta} \propto j^\alpha, $$
so that
$$ j^\mu \propto \partial_\tau f^{\mu \tau} + \partial _\nu f^{\mu
\nu}.$$
  Integrating this relation over $\tau$, with
vanishing boundary condition on the field strengths at $\tau
\rightarrow \pm \infty$ for any finite spacetime $x$, we see that the
$\tau$-integral of the fields satisfy Maxwell equations, with the
Maxwell current as the source.  We therefore identify the Maxwell
fields with the zero modes of what we shall call the pre-Maxwell
fields, i.e.,
$$ A^\mu(x) = \int_{-\infty}^\infty d\tau \, a^\mu
(x,\tau). \eqno(2.11)$$
Defining the Fourier representation
$$ a^\mu (x,\tau) = \int ds e^{is\tau} {\hat a}^\mu (x,s),
\eqno(2.12)$$
we see that the Maxwell limit corresponds to a field ${\hat
a}^\mu(x,s)$ with support in a small interval $\Delta s$ around zero,
so that
$$ a^\mu(x,\tau) \sim \Delta s {\hat a}^\mu(x,0). \eqno(2.13)$$
In the Maxwell theory, the field $A^\mu(x)$ must have dimension
($\ell$ is length) $\ell^{-1}$, and
therefore the dimension of $a^\mu$ must be $\ell^{-2}$,
and that of $e_0$, $\ell$.  Then the dimension of $f^{\alpha \beta}$
is $\ell^{-3}$, so that the second order form occurring in the
Lagrangian has dimension $\ell^{-6}$.  The action integral provides a
factor $d\tau d^4 x$, of dimension $\ell^5$, and therefore there must
be a coefficient of dimension $\ell$, which we shall call $\lambda$,
 for the quadratic term in field strengths.  It then follows that in
the Maxwell limit,
$$ e_0 a^\mu(x,\tau) \sim e_0 \Delta s A^\mu(x),$$
so that this width sets the scale for the
Maxwell limit of the theory, i.e., $\Delta s e_0$ corresponds to the
Maxwell electric charge.  We shall see below that it must coincide
with $1 \over \lambda$, the dimensional parameter
 introduced in the Lagrangian\footnote{*}{In a more complete theory,
this parameter should therefore emerge from a dynamical condition.}.
\par The action for the quantized fields is then (see ref. 11 for
further discussion, and a review of the standard Maxwell case by this method)
 ($ d^5 x \equiv d\tau d^4x$)
$$\eqalign{ S &= \int d^5x \bigl\{ -{\lambda \over 4} f^{\alpha \beta}
f_{\alpha \beta} + {i \over 2}\{\psi^\dagger {\partial \psi \over
 \partial \tau} - {\partial \psi^\dagger \over \partial \tau} \psi
\}\cr &-{1 \over 2M} \psi^\dagger (\partial^\mu
-ie_0a^mu)(\partial_\mu - ie_0a_\mu) \psi \cr
&\qquad + e_0 \psi^\dagger a_5 \psi -G \partial_\alpha a^\alpha +
{1 \over 2\lambda} G^2 \bigr\}, \cr} \eqno(2.14)$$
where we have omitted explicit $x,\tau$ dependences; the
operator-valued function $G(x)$ is a ghost field, providing a
canonical conjugate to $a_5$ (as for $A_0$ in the usual Maxwell
action).
\par The classical field equations corresponding to this action, in
the absence of the ghost field, take the form
$$ \lambda \partial_\alpha f^{\alpha \beta} = e_0 j^\beta, \eqno(2.15)$$
so that we see that $e_0/\lambda$ is to be identified
 with the Maxwell electric charge, and hence in the Maxwell limit, the
inverse correlation length $\Delta s \sim 1/\lambda$.  It would be of
interest to study the Ward identities of this theory, and establish
the relation of this structure to charge renormalization.
\par We further note from $(2.15)$ that, using a generalization of the
Lorentz gauge, $\partial_\alpha a^\alpha = 0$, the equation for the
source-free case becomes a d'Alembert equation with an additonal
second derivative with respect to $\tau$.  The differential operator
has the form
$$ \partial_\tau \partial^\tau -\partial_t^2 + \Delta = \sigma
\partial \tau^2 -\partial_t^2 + \Delta, $$
where $\sigma = \pm$ is the signature of the $\tau$ variable in the
wave equation ($\Delta$ is the Laplacian). Just as for the emergence
of electromagetism formally
from the non-relativistic Schr\"odinger theory, the evolution
parameter enters the manifold of the resulting wave equation. Positive
signature, corresponding to ${\rm O}(4,1)$ invariance of the
homogeneous equations, corresponds to a field with real mass $s$ under
Fourier transform, and with negative signature, to a tachyonic wave
equation, with ${\rm O}(3,2)$ invariance. The Green's functions for
these wave equations have been worked out.$^{16}$.  The integral over
$\tau$ for the tachyonic part of the Green's function vanishes, so
there is no violation of causality in the transmission of information
defined through Maxwell fields. The theory is, however, capable of
establishing dynamical correlations which are spacelike.
\par The canonical conjugate momenta are defined as
$$\eqalign{ \pi^\mu &= {\delta {\cal L} \over \delta(\partial_\tau
a_\mu)} = -\lambda f^{5\mu} \cr
\pi^5 &= {\delta {\cal L} \over \delta (\partial_\tau a_5)} = -\sigma
G(x) \cr
\pi_\psi &= {\delta {\cal L} \over \delta (\partial_\tau \psi)} =
i\psi^\dagger .\cr} \eqno(2.16)$$.
\par The Gauss law obtained from the classical equations $(2.15)$ is
$$ \lambda \partial_\mu f^{5\mu} = e_0 \rho, \eqno(2.17)$$
so that we should impose on physical states that
$$ \langle \partial_\mu \pi^\mu + j^5 \rangle = 0,$$
i.e., that the Gauss law holds.  This requirement can be satisfied by
imposing $G^{(+)}\vert \nu \rangle = 0$ (note that $G(x)$
 satisfies $(\sigma \partial_\tau^2 - \partial_t^2 + \Delta)G=0$,
so that it is a free field, and the $\pm$ frequency parts can be
isolated). The stability of this condition requires that
$$ {\dot G} ^{(+)}\vert \nu \rangle = 0 .\eqno(2.18)$$
We shall see that this condition is satisfied if the Gauss law is
true, so that the theory is consistent.
\par To generate the time evolution, we carry out the Legendre
transform of the Lagrangian to obtain the Hamiltonian of the system,
with the result that
 $$  K = K_\gamma + K_m + K_{\gamma m}, \eqno(2.19)$$
where
$$\eqalign{ K_\gamma &= \int d^4 x
 \bigl\{ -{1 \over 2\lambda}\pi^\mu
\pi_\mu - {\sigma \lambda \over 4} f^{\mu \nu}f_{\mu \nu} \cr
&\qquad + \pi^\mu (\partial_\mu a^5) -
\pi^5 (\partial_\mu a^\mu) - {\sigma
\over 2 \lambda} (\pi^5)^2 \bigr\} ,\cr
K_m &= {\sigma \over 2M} \int d^4 x \psi^\dagger \partial_\mu
\partial^\mu \psi ,\cr}$$
and
$$\eqalign{K_{\gamma m} &= \sigma \int
 d^4 x\bigl\{ -e_0\psi^\dagger a_5 \psi -
{ie_0 \over 2M} \psi^\dagger \lbrack 2a^\mu a_\mu +
\partial_\mu a ^\mu \rbrack \psi
\cr
&\qquad -{e_0^2 \over 2M} \psi^\dagger \psi
a^\mu a_\mu \bigr\}. \cr } $$
\par The canonical commutation relations are
$$ \eqalign{\lbrack \pi^\alpha (x), a_\beta (y)\rbrack
 &= -i\delta_\beta^\alpha
\delta(x-y)\cr
\lbrack i\psi^\dagger(x), \psi(y)\rbrack
&= -i\delta (x-y). \cr} \eqno(2.20)$$
 It then follows that
$$ {\dot G} = i\lbrack K,G\rbrack
= -\sigma(\rho + \partial_\mu\pi^\mu),
\eqno(2.21)$$
and hence the stability of the condition $(2.18)$ implies the Gauss
law.  The condition that the commutator of
${\dot G}$ with the Hamiltonian vanish in physical states, i.e., the
stability of Gauss law, is satisfied as well; one uses, after taking
the commutator of $\partial_\mu \pi^\mu$ with $K$, the current
conservation law $(2.7)$. The operator
$$e^{i\chi} = \exp {i\int d^4x \Lambda(x)
(\partial_\mu \pi^\mu(x) +
\rho(x))} \eqno(2.22)$$
commutes with the Hamiltonian in physical states, and can be used to
generate a gauge transformation which eliminates the part of the
$a^\mu$ field which is parallel to $\partial^\mu \Lambda$ in the
field functional $\Psi(a^\mu_\perp , a^\mu_\parallel , a^5, \psi)$,
i.e., the part $a^\mu_\parallel$.  Consider the following three
cases.$^{11}$

\noindent {\it Case 1\/}: $k^\mu$ timelike.

\noindent In this case, the component $a^0$ can be eliminated in the
frame $k^\mu = (k^0, 0,0,0),$ leaving the ``Coulomb'' potential $a^5$
and three polarizations $a_i$.  The polarization space is then
a positive norm representation of ${\rm O}(3)$.

\noindent {\it Case 2\/}: $k^\mu$ spacelike.

\noindent In the frame $k^\mu = (0,0,0,k^3)$, one can eliminate $a^3$,
and leaving the components $a^0, a^1, a^2 $.  These directions span
the indefinite space representing ${\rm O}(2,1)$.  The Casimir
operator $N=M_{12}^2 - M_{01}^2 - M_{02}^2$  is invariant (under the
dynamical action of the Hamiltonian, which is ${\rm O}(3,1)$
invariant), and the sign of its expectation value should therefore be
preserved under evolution.  The states of polarization with negative
norm removed is therefore a stable invariant subspace.  The zero-norm
components can be removed as in Case 3.

\noindent {\it Case 3\/}: $k^\mu$ lightlike.

\noindent In this case, one can eliminate $a_0$ and $a_\parallel$
together, leaving only two transverse polarizations, as in the Maxwell
theory.  The zero norm states are eliminated by means of
the Gupta-Bleuler condition, as is well known in the usual
 electromagnetic theory.
\par We see that the off-shell photons, which are massive (or
tachyonic), have three polarization degrees of freedom.  It is
therefore important to prove that for the equilibrium black-body
radiation field, which shows a specific heat characteristic of just
two polarization states, that the off-shell photons do not
contribute.  Although it could be expected that off-shell photons are
important at the walls, where emission and absorption take place, the
volume contribution  to the specific heat would not show this effect,
but careful estimates must be carried out.  There are many other
places where observable phenomena might exist, such as deep inelastic
scattering experiments, and these will be discussed elsewhere.
\par Carrying out the transformation $(2.22)$ on the Hamiltonian, only
polarization degrees of freedom remain$^{11}$, as for the usual Maxwell
case.  There is, as in the usual theory, where it emerges as an
instantaneous Coulomb interaciton, an additional residual term of the
form
$$ \eqalign{\langle K_c \rangle &= \bigl\langle -{1 \over 2\lambda} \int
d^4x\,d^4y \, \partial_x^\mu \pi_\mu(x) G(x-y) \partial_y^\nu
\pi_\nu(y) \bigr\rangle \cr
&= {e_0^2 \over 2 \lambda} \int d^4x\,d^4y \, \langle \rho_\tau(x)
G(x-y) \rho_\tau(y) \rangle + {\rm const}, \cr} \eqno(2.23)$$
 where we have used the Gauss law, and $G(x-y)$ is a Green's funciton
for the d'Alembert operator.
\par We have discussed above an interpretation linking the parameter
$\lambda$ with a correlation length of the fields in the Maxwell
limit.  Taking this correlation into acount, it was argued in ref. 11
that, in the classical limit, when the relative motion of the
particles is not too large, $(2.23)$ becomes equivalent to the Fokker
action$^{17}$.
\bigskip
\noindent
{\bf 3. Conclusions and Remarks.}
\smallskip
\par I have reviewed the second quantization of the gauge fields
generated by the \hfil\break
Stueckelberg-Schr\"odinger equation, and shown how
the resulting Hamiltonian can be represented in terms of the
polarization (physical) fields up to an additional term which is
approximately related to the Fokker action.  Just as in the usual
case, where the additional term represents the instantaneus Coulomb
field, our construction$^{11}$ provides an extra term that correponds
to a ``pre-theory'', i.e., a theory that could be effective before
quantum effects become important.
\par The quantum matter fields satisfy the (equal $\tau$) canonical
 commutation relations
  $$ \lbrack \psi_\tau (x), \psi_\tau ^\dagger(y)\rbrack =
\delta^4(x-y)\eqno(3.1)$$
Evidently, these fields generate a Fock space of a different type than
that generated by on-shell fields associate with a Klein-Gordon
equation. They create and annihilate particles of arbitrary mass; the
quantum states are contructed over wave functions which restrict these
masses according to the dynamical equations of the system.  In the
limit in which we may think of restricitng these masses to definite
values, it is of interest to see how this Fock space deforms to the
usual one.  To see this, let us take the Fourier transform, and
express $(3.1)$ in the form
$$ \lbrack \psi_\tau (p), \psi_\tau ^\dagger(p')\rbrack
= \delta^4(p-p')
\eqno(3.2)$$
Integrating both sides over $E$, the right side becomes $\delta^3({\bf
p} - {\bf p}')$; to carry out the integral over the left side, we use
the fact that the momentum must remain fixed in the differentiation,
and therefore only the mass can be varied (in the formula $E=
\sqrt{{\bf p}^2 + m^2}$), and hence
 $$dE = {1 \over 2E} dm^2. \eqno(3.3)$$
 If we call
$$  {\hat \psi}({\bf p}) = \sqrt{dm^2} \psi (p) \vert_{E=\sqrt{{\bf
p}^2 + m^2}} \eqno(3.4)$$
we see that the resulting commutation relations are
 $$ \lbrack {\hat \psi}_\tau({\bf p}),
  {\hat \psi}^\dagger_\tau({\bf p'})\rbrack =
2E\delta^3({\bf p} - {\bf p}').\eqno(3.6)$$
\par In a similar way, the space-time commutation relations $(3.1)$
vanish for $x^0 \neq y^0$.  Integrating over an infinitesimal interval
$dx^0$, the right hand side becomes $\delta^3({\bf x} - {\bf y})$,
and we may absorb factors $\sqrt{dx^0}$ into each space-time field
at equal time.  We therefore recover the usual equal time commutation
relations. A more complete discussion of the equal time limit of
the on mass shell theory will be given elsewhere.

\bigskip
\frenchspacing
\noindent
{\bf References}
\item{1.} E.C.G. Stueckelberg,  Helv. Phys. Acta {\bf 14},
372,588(1941); {\bf 15}, 23 (1942).
\item{2.} L.P. Horwitz and C. Piron, Helv. Phys. Acta {\bf 48}, 316
(1974).
\item{3.}R.E. Collins and J.R. Fanchi, Nuovo Cimento {\bf 48A}, 314
(1978); J.R. Fanchi and R.E. Collins, Found. Phys. {\bf 8}, 851 (1978).
\item{4.}J.R. Fanchi, {\it Parametrized Relativistic Quantum
 Theory}, Kluwer, Norwell, Mass. (1993).
\item{5.} J.R. Fanchi, Found. of Phys. {\bf 11}, 493
(1981);B. Thaller, J. Phys. A:Math. Gen {\bf 14}, 3067 (1981).
\item{6.} T.D. Newton and E.P. Wigner, Rev. Mod. Phys. {\bf 21}, 400
(1949).
\item{7.} L. Landau and R. Peierls, Zeits. f. Physik {\bf 69}, 56
(1931).
\item{8.} R. Arshansky and L.P. Horwitz, Found. Phys. {\bf 15}, 701
(1985).
\item{9.} R.I. Arshansky and L.P. Horwitz, Jour. Math. Phys. {\bf 30},
66,380 (1989).
\item{10.} K. Haller, Phys. Rev. D {\bf 36}, 1830 (1987), and
references therein; M. Henneaux and C. Teitelboim, {\it Quantization
of Gauge Systems\/},Princeton University Press, Princeton (1992).
\item{11.} N. Shnerb and L.P. Horwitz, Phys. Rev A {\bf 48}, 4068 (1993).
\item{12.} J. Schwinger, Phys. Rev. {\bf 74}, 1439 (1948);
S. Tomonaga, Prog. Theor. Phys. {\bf 1}, 27 (1946). See  H. Umezawa,
{\it Quantum Field Theory}, North Holland, New York (1956) for
discussion.
\item{13.} D. Saad, L.P. Horwitz and R.I. Arshansky, Found. Phys. {\bf
19}, 1126 (1989).
\item{13.} J.D. Jackson, {\it Classical Electrodynamics},2nd Edition,
 Wiley, New York (1975).
\item{15.} F. Rohrlich, {\it Classical Charged Particles\/},
Addison Wesley, Redwood City (1965) and (1990).
\item{16.} M.C. Land and L.P. Horwitz, Found. Phys.
 {\bf 21}, 299 (1991).
\item{17.} J.A. Wheeler and R.P. Feynman, Phys. Rev. {\bf 21},
 425 (1949).  See also R.P. Feynman, Phys. Rev. {\bf 80}, 440 (1950);
J. Schwinger, Phys. Rev. {\bf 82}, 664 (1951).

\vfill
\end
\bye